\documentclass{mem}
\usepackage{natbib}\usepackage{txfonts}\usepackage{balance}
\usepackage{graphicx}
\usepackage[a4paper,breaklinks,dvipdfm]{hyperref}
\idline{1}{1}
\begin{document}
\def\teff{$T\rm_{eff }$}
\def\kms{$\mathrm {km s}^{-1}$}

\title{
Ages and age spreads in young stellar clusters
}

   \subtitle{}

\author{
R.~D. Jeffries\inst{1} 
          }

\institute{
Astrophysics Group, Keele University, Keele, Staffordshire, ST5 5BG, UK
\email{r.d.jeffries@keele.ac.uk}
}

\authorrunning{Jeffries}

\titlerunning{Ages and age spreads}

\abstract{I review progress towards understanding the timescales of
  star and cluster formation and of the absolute ages of young stars. I
  focus in particular on the areas in which Francesco Palla made highly
  significant contributions -- interpretation of the
  Hertzsprung-Russell diagrams of young clusters and the role of
  photospheric lithium as an age diagnostic.
}
\maketitle{}

\section{Introduction}

Estimating the absolute ages of young stars and
ascertaining the extent of age spreads in young clusters is crucial in
understanding the mechanisms and timescales upon which stars form, 
upon which circumstellar disks disperse and planetary
systems assemble, and for understanding the role of varying stellar birth
environments on these issues. Francesco Palla produced 
highly influential work in these areas; my review focuses on
two key aspects: (i) the interpretation of the Hertzsprung-Russell
diagrams (HRDs) and colour-magnitude diagrams (CMDs) of young clusters
and star forming regions \citep{palla99, palla00, palla02}, and (ii) 
the use of photospheric lithium
abundance measurements as an orthogonal method to estimate and
calibrate young stellar ages \citep{palla05, palla07, sacco07}.

\section{Ages from H-R diagrams}

Low-mass stars ($\leq 2M_{\odot}$) take significant time
($\sim 10$--200\,Myr) to evolve from newly revealed T-Tauri stars to the
zero-age-main-sequence (ZAMS). This pre-main-sequence (PMS) evolution
occurs on mass-dependent timescales (faster for higher mass
stars); stars initially descend fully convective Hayashi tracks
followed by, for higher mass objects ($\geq 0.4M_{\odot}$), the
development of radiative cores and a blueward traverse along the Henyey
track before settling onto the ZAMS \citep[e.g.][]{iben65}.  In
principle, the construction of grids of mass-dependent evolutionary
tracks and corresponding isochrones in the HRD can be used with estimates of
luminosity and effective temperature ($T_{\rm eff}$) or equivalently (given
appropriate bolometric corrections), absolute magnitude and colour, to
yield ages and masses for PMS stars. An advantage to using low-mass
stars when studying young clusters, 
rather than their higher mass siblings, is they are much
more populous, allowing statistical analyses, and their movement in the
HRD can be much larger for a given age change.

In a series of papers, Francesco (and Steven Stahler) noted that, when
plotted on the HRD, stars are dispersed around the single
isochrones predicted by PMS
models. This indicated a substantial age spread of at least a few Myr,
and in some cases $>10$ Myr. The pattern was repeated in several 
young clusters and when ages inferred from HRD position
were turned into a star formation history, suggested an accelerating
star formation rate as a function of (linear) time. This highly-cited
result has launched a thousand telescope proposals and
is still hotly debated. Does an extended star formation history
indicate inefficient star formation moderated by turbulence and magnetic
fields, or can the spreads be explained by observational uncertainties
and problems with PMS models so that actually, star formation is rapid
and efficient, taking place on dynamical timescales?

Opponents of the idea of large age spreads have pointed to the role of
astrophysical effects and observational uncertainties in scattering
stars in the HRD, giving the {\it impression} of a large age
dispersion. \citet{hartmann01} noted that the apparent age distribution
was lognormal, with $\sigma \sim 0.4$~dex, perhaps reflecting the logarithmic nature of
uncertainties in luminosity estimates and that age $\propto L^{-3/2}$
on Hayashi tracks. There are uncertainties in distance,
extinction, and also due to intrinsic variability, accretion and the
presence of binaries that must certainly be accounted for in
estimating a true age dispersion. Detailed simulations by
\citet{reggiani11} and \citet{preibisch12} concluded that whilst
these effects were important, they probably do not explain the entire
extent of observed dispersions. 

It seems certain
that the very old ages assigned to at least 
some PMS stars in young clusters are
due to mis-estimated luminosities and temperatures associated
with an incorrect or at least incomplete treatment of extinction and
accretion \citep{manara13}. On the other hand, 
support for genuine dispersions in
luminosity (or radius at a given $T_{\rm eff}$) has 
been found by considering the distribution
of projected radii (rotation period multiplied by projected rotation
velocity) in the Orion Nebula cluster (ONC) and IC~348
\citep{jeffries07, cottaar14} and from the IN-SYNC APOGEE survey that
finds a significant correlation between increasing age and
spectroscopic gravity in the same clusters \citep{cottaar14,
  dario16}. 

There seems little doubt that a fraction of the observed age
dispersion must be due to sources of astrophysical and observational
uncertainty, but also strong evidence that at least some of the luminosity
and radius spread is real. Whether this implies
genuine age spreads requires evidence from other observations
and independent astrophysics.

\section{Lithium as an age indicator}

Lithium is ephemeral in low-mass stellar photospheres. As PMS stars
contract, their cores reach Li-burning temperatures before reaching the ZAMS. If
the convection zone base is also above the Li-burning
temperature (which it would be in fully convective stars)
then photospheric Li is also depleted on timescales less than a few
Myr. The age at which core Li burning begins is mass-dependent (later
for lower mass stars), but the development of a radiative core can
arrest photospheric Li depletion in more massive objects. These
phenomena lead to a complex, but age-dependent, behaviour for
Li abundance as a function of luminosity, $T_{\rm eff}$ or colour.

Palla et al. (2005, 2007) were among the first to suggest Li depletion
could serve as an independent test of
ages in very young low-mass stars. Li depletion is expected 
to begin in stars of $\sim
0.5M_{\odot}$ at an age of about 5\,Myr and subsequently develops at higher and
lower masses. Since the physics of Li depletion is comparatively simple,
it has been argued that this currently provides the {\it least}
model-dependent means of estimating young stellar ages
\citep[e.g.][]{soderblom14}, however masses cannot be measured directly
so one relies on colours, $T_{\rm eff}$ or (better) luminosities
as proxies.

Palla et al. (2005) and Sacco et al. (2007) found examples of
Li-depleted low-mass stars that appeared older than 10\,Myr in the ONC
and the $\sigma$ Ori cluster, and much older
than the bulk of their siblings, perhaps supporting the notion of large
age spreads $>10$ Myr. Subsequent work by
\citet{sergison13} on the ONC and NGC~2264 confirmed the
presence of a dispersion in Li abundance, but noted the difficulty in
assessing Li abundances for PMS stars that are often accreting. Any
veiling continuum weakens the Li\,{\sc i}~6708\AA\ line that is
exclusively used; this
combined with the saturated nature of this strong resonance line can
lead to the mistaken inference of significant Li
depletion. \citet{lim16} took the expedient option of excluding stars
with signs of accretion from their analysis (which one might presume
were younger stars), still finding evidence for some age
dispersion in NGC~2264, but with an absolute value $\leq 4$ Myr and smaller
than the spread implied by the HRD.

Taken at face value, the combined information from Li depletion,
the HRD and spectroscopic indicators of radii suggests that some
dispersion in age is present, but probably no more than a few Myr and
not as much as suggested by the HRD alone. However, there are problems
that have emerged even with this simple interpretation that may betray
interesting facets of PMS evolution that have yet to be correctly incorporated.

\section{Problems with evolutionary models}

\begin{figure*}[t!]
\resizebox{\hsize}{!}{\includegraphics[clip=true]{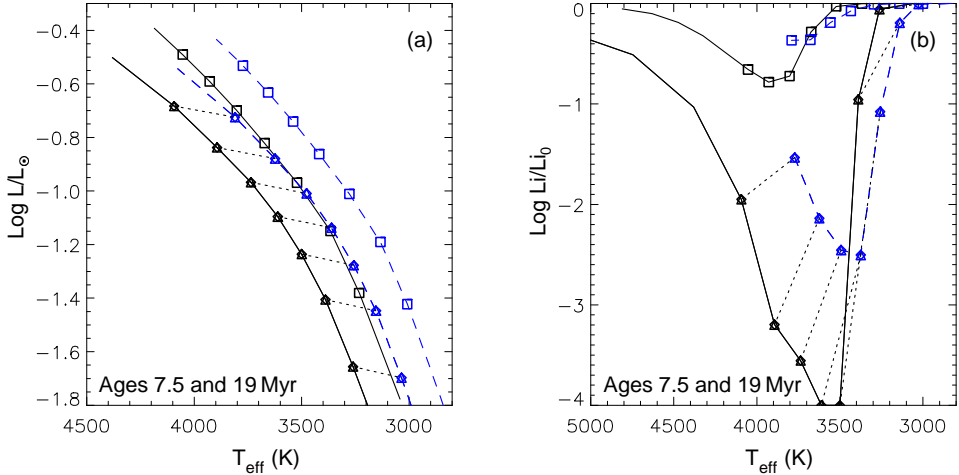}}
\caption{\footnotesize
The effects of 10\% inflation due to magnetic activity. The left hand
panel shows isochrones in the HRD from Baraffe et al. (2015) at 7.5
(diamonds) and 19 Myr (squares). The dashed lines show the same
isochrones modified for the effects of radius inflation. Mass points
from $0.2M_{\odot}$ to $0.8M_{\odot}$ in $0.1M_{\odot}$ steps are
indicated by open symbols on each isochrone. The dotted lines indicate
the movement of a star of a given mass due to radius inflation. Note
how an inflated 19 Myr isochrone lies almost on top of the 7.5 Myr
uninflated isochrone. The right hand panel is similar but shows the
effects of radius inflation on the expected level of Li
depletion. These diagrams (adapted from Jeffries et al. 2017 by
R. Jackson priv. comm.)
illustrate that radius inflation acts to reduce luminosity, lower
$T_{\rm eff}$ and decrease Li depletion for a
star of a given mass and age.
}
\label{fig1}
\end{figure*}

(i) {\it Why is Li depletion correlated with rotation?} That
rapidly rotating low-mass stars appear to preserve their Li longer, has been established
in older clusters and becomes clearer with better data
\citep{barrado16}. This trend is now becoming apparent at
even younger ages and may be responsible for some of the Li depletion
dispersion previously claimed to be associated with an age spread
\citep{bouvier16}. Since PMS stars are expected to spin-up as they
contract, then older stars ought to be faster rotating and {\it more} Li
depleted if the age dispersion were genuine.

(ii) {\it Why do
  Li-depletion ages disagree with isochronal ages from the HRD?}
\citet{jeffries17} have pointed out that Li depletion ages
and HRD/CMD ages are not completely independent; Li depletion takes
places when the core temperature, and hence mass to radius ratio,
reaches a certain threshold, whilst HRD/CMD ages also depend on radius at a given
$T_{\rm eff}$, though not as sensitively. The {\it same} evolutionary
models give significantly younger ages for low-mass PMS stars in the
$\gamma^2$ Velorum cluster than implied by the strong Li depletion
seen in its M-dwarfs. The Li depletion also takes place at much redder
colours and lower inferred $T_{\rm eff}$ than expected. 
The CMD and Li-depletion pattern cannot
be explained simultaneously by any commonly used evolutionary
codes at any age.

(iii) {\it Why are more massive stars in young clusters
  judged to be older than the low-mass stars?} The ages of
clusters with PMS stars can also be estimated by looking at how far from the
ZAMS towards the terminal-age main-sequence their high mass ($>5M_{\odot}$)
stars have progressed. When done with a self-consistent and accurate
treatment of reddening \citet{naylor09} suggested that the high-mass
stars were significantly older than their low-mass
siblings by a factor of two. 
This was followed-up with a larger
sample by \citet{bell13}, who demonstrated that the
low-mass ages could be brought into agreement with the high-mass ages
(and ages from Li depletion)
with systematic changes in the bolometric corrections adopted by the
models.

(iv) {\it Why do current models fail to correctly predict the
  location of PMS eclipsing binary components in the HRD?} New
examples found in star forming regions provide
challenges to evolutionary models. Their masses and radii are not
well predicted from their estimated luminosities and $T_{\rm eff}$
\citep{kraus15, david16}. The PMS binary components appear colder than
predicted by the models and more luminous than predicted at the age of
higher mass stars in the same clusters.

These problems have lead to consideration of whether PMS evolutionary
models are yielding the correct absolute masses, ages and hence age spreads at all.
An idea that has gained some traction is that episodic accretion
during the first million years of a star's life can significantly
influence both the HRD position and Li depletion
\citep{baraffe10, baraffe17}. Variations in
accretion rate and the exact timing of accretion could lead to
apparent age dispersions and to the occasional star
appearing much older in the HRD and/or exhibiting significant Li
depletion.

An alternative that is also gaining support is that magnetic activity
may ``inflate'' low-mass stars (or at least slow their contraction),
either through magnetic inhibition of convection \citep{mullan01,
  feiden14} or the blocking of radiative flux by cool starspots
\citep{jackson14, somers15b}. These ideas have the attraction that we
know young low-mass stars are magnetically active and that they have
extensive starspot coverage \citep[some recent spectroscopic estimates
  suggest more than 50\%,][]{gully17}.

Let us suppose then that active low-mass PMS stars are inflated by
$\sim 10$\% compared to the predictions of ``standard'' evolutionary
models at a given mass and age. This is roughly the level suggested by
recent modelling work that attempts to incorporate the effects of
suppressed convection or starspots.  Jeffries et al. (2017) \citep[see
  also][]{feiden16, messina16} have shown that such stars become cooler
and only slightly less luminous. The net result is that stars move
almost horizontally in the HRD resulting in severely underestimated ages and
masses when using ``standard'' models (see Fig.~1). At the same time their core
temperatures are reduced, delaying the onset of Li depletion and
decreasing the $T_{\rm eff}$ of stars in which Li depletion is first
seen.

If magnetic models such as those of Jackson \& Jeffries (2014),
\citet{somers15a} or Feiden (2016) are adopted, then HRD/CMD ages are
brought into much closer agreement with the Li depletion ages, but at
the expense of {\it doubling} the ages inferred from the HRD (see Fig.~1). This also
brings ages from low-mass and high-mass stars into broad agreement,
potentially solves the problems with eclipsing binary parameters
\citep{macdonald17} and
could introduce a dispersion into the HRD and Li-depletion patterns of
young stars that is correlated with rotation and/or magnetic activity \citep{somers15a}.
If correct, such a large shift has considerable implications for the
timescales of PMS evolution, the dispersal of circumstellar matter and
hence the time available to form planetary systems, all of which are
keyed-in to the absolute timescales set by age estimates for young,
low-mass stars.

\section{Summary}

The investigation of ages and age spreads in young clusters using the
HRD and Li-depletion, begun by Francesco Palla and colleagues, remains a
vibrant and controversial topic. Current evidence suggests that age
spreads are a lot smaller than 10\,Myr (within a single cluster), but
that not all the dispersion in cluster HRD/CMDs and Li depletion can be
explained by observational and astrophysical uncertainties. Some of the
observed spread does appear to be due to a genuine distribution of
radius among stars with similar $T_{\rm eff}$ and mass, which might be
attributable to a modest age spread of a few Myr.
We are now moving into an era of more sophisticated stellar modelling
that questions the veracity of both the absolute ages of PMS stars and
the inferred age spreads in young star forming regions.

\bibliographystyle{aa}

%\bibliography{references}

\end{document}